\documentstyle[12pt]{article}
\makeatletter
\@addtoreset{equation}{subsection}
\makeatother

\begin{document}
\begin{center}
{\Large \bf
Indeterministic Quantum Gravity} \\[0.5cm]
{\large\bf V. Dynamics and Arrow of Time} \\[1.5cm]
{\bf Vladimir S.~MASHKEVICH}\footnote {E-mail:
mashkevich@gluk.apc.org}  \\[1.4cm]
{\it Institute of Physics, National academy
of sciences of Ukraine \\
252028 Kiev, Ukraine} \\[1.4cm]
\vskip 1cm

{\large \bf Abstract}
\end{center}

This paper is a continuation of the papers [1-4] and is
devoted to the riddle of the origin of the arrow of time.
The problem of time orientation reduces to that of the
difference between the past and the future. The riddle
escapes solution in deterministic dynamics and in the dynamics
of standard indeterministic quantum theory as well. In the
dynamics of indeterministic quantum gravity, the past is
reconstructible uniquely, whereas the future may be
forecasted only on a probabilistic level. Thus the problems
of the past and the future and, by the same token, of time
orientation are solved.

\newpage

\hspace*{6 cm}
\begin{minipage} [b] {9 cm}
Time present and time past

Are both perhaps present in time future,

And time future contained in time past.
\end{minipage}
\begin{flushright}
Thomas Sterns Eliot \vspace*{0.8 cm}
\end{flushright}

\begin{flushleft}
\hspace*{0.5 cm} {\Large \bf Introduction}
\end{flushleft}

One of the most ancient riddles of physics is that of the
origin of the arrow of time, or of the nature of the
difference between the past and the future. It is conventional
to search for a solution to this problem in dynamics, i.e.,
time evolution of a state of a physical system. The solution
may be given by a dynamics which is asymmetric, or orientable in
the sense of the sequence of states.

Deterministic dynamics does not involve such an orientability.
There exists an established opinion that in standard
indeterministic quantum dynamics there is no arrow of time
as well [5].

The dynamics of indeterministic quantum gravity---of the theory
being developed in this series of papers---features an
asymmetry, which may be used for determining the arrow of
time.

This paper is dedicated to a comprehensive consideration of
the issues outlined above.

In Section 1, a general treatment of dynamics is given. A
predynamical time, or pretime, is introduced, and the problem
of the arrow of physical time consists in fixing the
direction of the latter with respect to the direction of the
former. Physical time is oriented from the past to the
future. The idea of defining these notions is that the
predeterminability of the future should be less than the
reconstructibility of the past.

In Section 2, the dynamics of standard indeterministic quantum
theory is examined from the standpoint of the general
treatment. This dynamics is symmetric and does not give rise
to a choice of the future and the past and, by the same
token, to physical time orientation.

In Section 3, a new scheme for quantum jumps in indeterministic
quantum gravity is introduced, and then the related dynamics
is analyzed. In this dynamics, the past is reconstructible
uniquely, whereas the future may be forecasted only on a
probabilistic level. Thus the problems of the future and the
past and, by the same token, of physical time orientation
are solved.

\section{Dynamics and time orientation}

We introduce a treatment of dynamics which may be readily
generalized for a subsequent application to cosmology.

\subsection{Predynamical and physical time}

Dynamics in general is time dependence of a state of a
physical system. A definition of the state may involve the
direction of time, which is not given a priori. Therefore we
introduce a predynamical time, or pretime, for short, $\tau$
as a point of the oriented real axis $T$. For the direction of
physical time, $t$, there are two possibilities: $t=\tau$
and $t=-\tau$. Definitions of the state and dynamics should
be given in terms of the pretime, and the problem of
physical time orientation is to be solved on the basis of
dynamics.

\subsection{Dynamical process}

We start with the notion of a dynamical process, which
plays a central role in dynamics. Let $\Omega$ be a set of
pure states $\omega$, $\Delta$ be a connected subset of
$T$, i.e., an interval:
\begin{equation}
\Delta=(\tau_{1},\tau_{2}),[\tau_{1},\tau_{2}),
(\tau_{1},\tau_{2}],[\tau_{1},\tau_{2}],\quad
-\infty\leq \tau_{1}<\tau_{2}\leq\infty.
\label{1.2.1}
\end{equation}
A dynamical process ${\cal P}_{\Delta}$ on ${\Delta}$
is a function
from $\Delta$ to $\Omega$:\begin{equation}
{\cal P}_{\Delta}:\Delta\rightarrow \Omega,\quad \Delta\ni\tau
\mapsto {\cal P}_{\Delta}(\tau)=\omega_{\tau}\in\Omega.
\label{1.2.2}
\end{equation}

In fact, it would suffice for ${\cal P}_{\Delta}(\tau)$
to be defined almost everywhere on $\Delta$.

A restriction and extension of a process are defined as
those of a function with regard to the fact that the
domain of the process is connected.

A left (right) prolongation of a process ${\cal P}_{\Delta}$ to
$\Delta'$, $\Delta'\ni\tau'<\tau\in\Delta$ ($\Delta\ni\tau
<\tau'\in\Delta'$), is a process ${\cal P}_{\Delta'}$, such that
there exists a process ${\cal P}_{\Delta\cup\Delta'}$ with the
restrictions ${\cal P}_{\Delta}$ and ${\cal P}_{\Delta'}$.

\subsection{Dynamics}

Dynamics on $\Delta$, ${\cal D}_{\Delta}$, is a family of
processes
on $\Delta$:
\begin{equation}
{\cal D}_{\Delta}=\left\{ {\cal P}_{\Delta} \right\}.
\label{1.3.1}
\end{equation}

A restriction and extension of a dynamics boil down to those
of corresponding processes.

\subsection{Deterministic process and deterministic
dynamics}

A deterministic process ${\cal P}_{\Delta}$ is defined as follows:
For every restriction of ${\cal P}_{\Delta}$ the only extension to
$\Delta$ is ${\cal P}_{\Delta}$ itself.

A deterministic dynamics is a family of deterministic
processes.

\subsection{Indeterministic point, process, and dynamics}

An interior isolated indeterministic point
$\tau\in{\rm int}\:\Delta$
of a process ${\cal P}_{\Delta}$ is defined as follows:
There
exists $\theta>0$, such that

(i) $(\tau-\theta,\tau+\theta)\subset\Delta$;

(ii) left prolongations of ${\cal P}_{\Delta}|_{[\tau,
\tau+\theta) }$
to ($\tau-\theta,\tau$) and right ones of ${\cal P}_{\Delta}|_
{(\tau-\theta,\tau]}$ to ($\tau,\tau+\theta$) are
deterministic processes;

(iii) cardinal numbers ${\rm card^{left}}$ and
${\rm card^{right}}$ of sets of those prolongations meet
the condition ${\rm card^{left}+card^{right}}>2$.

We assume that there are only isolated indeterministic
points.

An indeterministic process is that with indeterministic
points. An indeterministic dynamics is one with
indeterministic processes.

\subsection{Orientable dynamics and time orientation:
The future and the past}

Let $\tau\in{\rm int}\:\Delta$ be an indeterministic point
of a process ${\cal P}_{\Delta}$. We introduce five dynamics
related to the point as follows:

(i) ${\cal D}_{(\tau-\theta,\tau)}$, ${\cal D}_{(\tau,
\tau+\theta)}$ are deterministic,
${\cal D}_{(\tau-\theta,\tau)}\ni {\cal P}_{\Delta}|_{
(\tau-\theta,\tau)}$, ${\cal D}_{(\tau,\tau+\theta)}\ni
{\cal P}_{\Delta}|_{(\tau,\tau+\theta)}$;

(ii) ${\cal D}_{(\tau-\theta,\tau+\theta)}=\{{\cal P}
_{(\tau-\theta,\tau+\theta)}^{\alpha},\alpha\in {\cal A}\}$,

${\cal D}_{(\tau-\theta,\tau+\theta)}|_{(\tau-\theta,\tau)}=
{\cal D}_{(\tau-\theta,\tau)}$, ${\cal D}_{(\tau-\theta,\tau+
\theta)}|_{(\tau,\tau+\theta)}={\cal D}_{(\tau,\tau+\theta)}$;

(iii) a graph where points are elements of ${\cal D}_{(\tau-
\theta,\tau)}$ and ${\cal D}_{(\tau,\tau+\theta)}$ and lines
connecting related points are elements of ${\cal D}
_{(\tau-\theta,\tau+\theta)}$ is connected and complete, i.e.,
involves all processes associated with the indeterministic
point;

(iv) ${\cal D}_{(\tau-\theta,\tau+\theta)}^{{\rm right}\;
\alpha}=
\{{\cal P}^{\alpha'}\in{\cal D}_{(\tau-\theta,\tau+\theta)}:
{\cal P}^{\alpha'}|_{(\tau-\theta,\tau)}={\cal P}^{\alpha}|
_{(\tau-\theta,\tau)}\}$,

${\cal D}_{(\tau-\theta,\tau+\theta)}^{{\rm left}\;\alpha}=
\{{\cal P}^{\alpha'}\in{\cal D}_{(\tau-\theta,\tau+\theta)}:
{\cal P}^{\alpha'}|_{(\tau,\tau+\theta)}={\cal P}^{\alpha}|
_{(\tau,\tau+\theta)}\}$,

${\rm card^{right\;\alpha},\; card^{left\;\alpha}}$
being corresponding cardinal numbers.

An indeterministic point is symmetric (asymmetric) if
${\rm card^{right\;\alpha}=(\ne)card^{left\;\alpha}}$.
A symmetric dynamics is
that with symmetric indeterministic points only.

An indeterministic dynamics is orientable if for all
indeterministic points either
\begin{equation}
{\rm card^{right\;\alpha}}>
{\rm card^{left\;\alpha}}\quad{\rm and}\quad
{\rm card}\:D_{(\tau,\tau+\theta)}>
{\rm card}\:D_{(\tau-\theta,\tau)}
\label{1.6.1}
\end{equation}
or
\begin{equation}
{\rm card^{right\;\alpha}<card^{left\;\alpha}}\quad
{\rm and}\quad
{\rm card}\:D_{(\tau,\tau+\theta)}<
{\rm card}\:D_{(\tau-\theta,\tau)}.
\label{1.6.2}
\end{equation}
An orientable dynamics is oriented as follows: The future
corresponds to the greater of
${\rm card^{right\;\alpha}}$,
${\rm card}\:D_{(\tau,\tau+\theta)}$
and
${\rm card^{left\;\alpha}}$,
${\rm card}\:D_{(\tau-\theta,\tau)}$,
i.e.,
\begin{equation}
{\rm card^{future\;\alpha}>card^{past\;\alpha}},\quad
{\rm card}\:D_{\rm future}>{\rm card}\:D_{\rm past};
\label{1.6.3}
\end{equation}
so that the physical time is
\begin{equation}
t=+(-)\tau\quad {\rm for}\;{\rm card^{right\;\alpha}>(<)
\;card^{left\;\alpha}},\;
{\rm card}\:D_{(\tau,\tau+\theta)}>(<)\;
{\rm card}\:D_{(\tau-\theta,\tau)}.
\label{1.6.4}
\end{equation}
This defines time orientation, or the arrow of time.

\subsection{Nonpredeterminability and the question of
reconstructibility}

For an oriented dynamics we have
\begin{equation}
{\rm card^{future\;\alpha}+card^{past\;\alpha}}>2,
\label{1.7.1}
\end{equation}
\begin{equation}
{\rm card^{future\;\alpha}>card^{past\;\alpha}}\geq 1,
\label{1.7.2}
\end{equation}
so that
\begin{equation}
{\rm card^{future\;\alpha}}>1.
\label{1.7.3}
\end{equation}
This implies that the future is not predeterminate.

If
\begin{equation}
{\rm card^{past\;\alpha}}=1,
\label{1.7.4}
\end{equation}
the past is reconstructible.

In any case, the inequality (\ref{1.6.1}) implies that the
reconstructibility of the past is greater than the
predictability of the future. This feature is inherent in
an oriented dynamics.

The phenomenon of memory should be related to dynamics
orientation.

\subsection{Probabilistic dynamics}

Let $\tau$ be an indeterministic point of a process
${\cal P}_{\Delta}$.
Time evolution implies transitions from one of the sets
${\cal D}_{(\tau-\theta,\tau)}$, ${\cal D}_{(\tau,
\tau+\theta)}$
to the other: from ${\cal D}^{\rm initial}$ to ${\cal D}
^{\rm final}$.
We assume that for their cardinal numbers
\begin{equation}
{\rm card^{final}\geq card^{initial}}
\label{1.8.1}
\end{equation}
holds.

Let there exist $i\to f$ transition probabilities, or
conditional probabilities $w(f/i)$, where $i$ and $f$ are
indexes of elements of ${\cal D}^{\rm initial}$ and
${\cal D}^{\rm final}$
respectively. The probabilities meet the equation
\begin{equation}
\sum_{f}w(f/i)=1.
\label{1.8.2}
\end{equation}
Taking into account the relations
\begin{equation}
{\rm card^{initial}}=\sum_{i}1=\sum_{i}\sum_{f}w(f/i)=
\sum_{f}\sum_{i}w(f/i)\leq \sum_{f}1={\rm card^{final}},
\label{1.8.3}
\end{equation}
we put
\begin{equation}
\sum_{i}w(f/i)\leq 1.
\label{1.8.4}
\end{equation}
By Bayes formula, the a posteriori probability is
\begin{equation}
w(i/f)=\frac{w(i)w(f/i)}{\sum_{i'}w(i')w(f/i')}.
\label{1.8.5}
\end{equation}
We put for the a priori probability
\begin{equation}
w(i)={\rm const},
\label{1.8.6}
\end{equation}
then
\begin{equation}
w(i/f)=\frac{w(f/i)}{\sum_{i'}w(f/i')}\geq w(f/i).
\label{1.8.7}
\end{equation}
Thus
\begin{equation}
w(f/i)\leq w(i/f)\leq 1.
\label{1.8.8}
\end{equation}

For a symmetric dynamics,
\begin{equation}
{\rm card^{final}=card^{initial}},\quad \sum_{i}w(f/i)=1,
\label{1.8.9}
\end{equation}
so that
\begin{equation}
w(i/f)=w(f/i).
\label{1.8.10}
\end{equation}
Specifically, the increase of entropy is, on the average, the
same for the future and for the past:
\begin{equation}
-\sum_{f}w(f/i)\ln w(f/i)\approx-\sum_{i}w(i/f)\ln w(i/f).
\label{1.8.11}
\end{equation}

\subsection{Irreversibility and orientation}

It should be particularly emphasized that irreversibility
does not imply dynamics orientability and, by the same
token, time orientation.

Indeed, a reversible dynamics is defined as follows. Let
${\cal P}_{\Delta}$ be a process with a symmetric domain, i.e.,
$\Delta=(\tau_{1},\tau_{2})$ or $[\tau_{1},\tau_{2}]$.
The inverse process, ${\cal P}_{\Delta}^{\rm inv}$, is
defined by
\begin{equation}
{\cal P}_{\Delta}^{\rm inv}(\tau)={\cal P}_{\Delta}(\tau_{1}+
\tau_{2}-\tau),
\quad \tau\in\Delta.
\label{1.9.1}
\end{equation}
Let $S$ be a transformation of $\Omega$, $S:\Omega\to\Omega$.
The transformed process, $S{\cal P}_{\Delta}$, is defined by
\begin{equation}
S{\cal P}_{\Delta}(\tau)=S({\cal P}_{\Delta}(\tau)),\quad
\tau\in\Delta.
\label{1.9.2}
\end{equation}
A dynamics ${\cal D}_{\Delta'}$ is reversible if there exists
a bijection $S:\Omega\to\Omega$, such that
\begin{equation}
S\;{\rm is\;an\;involution}\;(S^{2}\;{\rm is\; identity)
\; and}\; {\cal P}
_{\Delta}\in {\cal D}_{\Delta}
\Rightarrow {\cal P}_{\Delta}^{\rm rev}\equiv S{\cal P}_
{\Delta}^{\rm inv}
\in {\cal D}_{\Delta}\; {\rm for\; all}\; \Delta\subset
\Delta'
\label{1.9.3}
\end{equation}
(rev stands for reverse).

The nonexistence of $S$ does not imply the orientability of
${\cal D}_{\Delta'}$.

Here is an example. Let a dynamical equation be of the form
\begin{equation}
\frac{d^{2}x}{d\tau^{2}}=-\alpha\frac{dx}{d\tau}.
\label{1.9.4}
\end{equation}
All dynamical processes ${\cal P}_{(-\infty,\infty)}$ are
given by
\begin{equation}
{\cal P}_{(-\infty,\infty)}(\tau)=\omega_{\tau}=\left( x(\tau),
\frac{dx(\tau)}{d\tau} \right),
\label{1.9.5}
\end{equation}
\begin{equation}
x(\tau)=c_{1}+c_{2}e^{-\alpha\tau};
\label{1.9.6}
\end{equation}
they are deterministic.
The dynamics ${\cal D}_{(-\infty,\infty)}$
is irreversible but deterministic and, therefore, not
orientable.

On the other hand, a dynamics with asymmetric indeterministic
points is irreversible---in view of inequality ${\rm card
^{right\:\alpha}\ne card^{left\:\alpha}}$. Specifically, an
orientable
dynamics is irreversible.

\section{Dynamics of standard indeterministic quantum
theory}

Let us consider the dynamics of standard, or orthodox
indeterministic quantum theory from the standpoint
developed in the previous section.

\subsection{Standard dynamical process}

In standard quantum theory, indeterminism originates from
quantum jumps. A standard indeterministic dynamical process
${\cal P}_{(-\infty,\infty)}$ may be described as follows. Let
$\tau_{k},\;k\in K=\{0,\pm 1,\pm 2,...\}$, be indeterministic
points, i.e., points of jumps. The process is denoted by
\begin{equation}
{\cal P}_{(-\infty,\infty)}^{\{j_{k},k\in K\}},\;j_{k}\in
J=\{1,2,...,
j_{\rm max}\},\;j_{\rm max}\leq \infty.
\label{2.1.1}
\end{equation}
The definition of this process reduces to that of its
restrictions to the intervals
\begin{equation}
\Delta_{k}=(\tau_{k},\tau_{k+1}),\quad k\in K,
\label{2.1.2}
\end{equation}
\begin{equation}
{\cal P}_{\Delta_{k}}^{j_{k}}\equiv {\cal P}_{(-\infty,
\infty)}^{\{j
_{k},k\in K\}}|_{\Delta_{k}}.
\label{2.1.3}
\end{equation}
The process (\ref{2.1.3}) is defined as follows:
\begin{equation}
{\cal P}_{\Delta_{k}}^{j_{k}}(\tau)=\omega^{j_{k}}_{\tau}=
(\Psi^{j_{k}}
(\tau),\cdot\Psi^{j_{k}}(\tau)),\quad \tau\in\Delta_{k},
\label{2.1.4}
\end{equation}
where $\Psi^{j_{k}}$ is a state vector,
\begin{equation}
\Psi^{j_{k}}(\tau)=U(\tau,\tau_{k})\Psi_{j_{k}},
\label{2.1.5}
\end{equation}
\begin{equation}
A_{k}\Psi_{j_{k}}=a_{j_{k}}\Psi_{j_{k}},
\label{2.1.6}
\end{equation}
where the $A_{k}$ is an observable,
and $U$ is a unitary operator
of time evolution.

This description seemingly fixes the time orientation, namely,
in view of eq.(\ref{2.1.5}),
\begin{equation}
t=\tau.
\label{2.1.7}
\end{equation}
But there is another possibility for describing the process
considered.

\subsection{Reverse description}

In place of eqs.(\ref{2.1.5}),(\ref{2.1.6}), we may put
\begin{equation}
\Psi^{j_{k}}(\tau)=U(\tau,\tau_{k+1})\Psi_{j_{k+1}}^{\rm rev},
\label{2.2.1}
\end{equation}
\begin{equation}
A_{k+1}^{\rm rev}\Psi_{j_{k+1}}^{\rm rev}=a_{j_{k+1}}
^{\rm rev}\Psi_{j_{k+1}}^{\rm rev},
\label{2.2.2}
\end{equation}
where
\begin{equation}
\Psi_{j_{k+1}}^{\rm rev}=U(\tau_{k+1},\tau_{k})\Psi_{j_{k}},
\label{2.2.3}
\end{equation}
\begin{equation}
A_{k+1}^{\rm rev}=U(\tau_{k+1},\tau_{k})A_{k}U(\tau_{k},
\tau_{k+1}),
\label{2.2.4}
\end{equation}
\begin{equation}
a_{j_{k+1}}^{\rm rev}=a_{j_{k}}.
\label{2.2.5}
\end{equation}

This description implies, in view of eq.(\ref{2.2.1}), the time
orientation
\begin{equation}
t=-\tau.
\label{2.2.6}
\end{equation}

The two descriptions are completely equivalent physically.

\subsection{Standard quantum dynamics}

We have for an indeterministic point $\tau_{k}$
\begin{equation}
{\cal D}_{k}^{\rm left}\equiv{\cal D}
_{(\tau_{k}-\theta,\tau_{k})}=
\{{\cal P}_{\Delta_{k-1}}^{j_{k-1}}|_{(\tau_{k}-\theta,
\tau_{k})}
,j_{k-1}\in J\},
{\cal D}_{k}^{\rm right}\equiv{\cal D}
_{(\tau_{k},\tau_{k}+\theta)}=
\{{\cal P}_{\Delta_{k}}^{j_{k}}|_{(\tau_{k},\tau_{k}+\theta)},
j_{k}\in J\},
\label{2.3.1}
\end{equation}
\begin{equation}
{\rm card}\:{\cal D}_{k}^{\rm left}={\rm card}\:{\cal D}_{k}
^{\rm right}=
{\rm card}\:J.
\label{2.3.2}
\end{equation}
Thus standard quantum dynamics is not orientable.

\subsection{Standard probabilistic quantum dynamics}

We have for the time orientation $t=\tau$
\begin{equation}
w(j_{k+1}/j_{k})=w_{j_{k+1}\gets j_{k}}=|(\Psi^{j_{k+1}}
(\tau_{k+1}+0),\Psi^{j_{k}}(\tau_{k+1}-0))|^{2}=
|(\Psi_{j_{k+1}},U(\tau_{k+1},\tau_{k})\Psi_{j_{k}})|^{2},
\label{2.4.1}
\end{equation}
\begin{equation}
w_{j_{k+m}\gets j_{k+m-1}\gets\ldots\gets j_{k+1}\gets j_{k}}
=w(j_{k+m}/j_{k+m-1})\cdots w(j_{k+1}/j_{k});
\label{2.4.2}
\end{equation}
for the time orientation $t=-\tau$
\begin{equation}
w(j_{k}/j_{k+1})=w_{j_{k}\gets j_{k+1}}=
|(\Psi^{j_{k}}(\tau_{k+1}-0),\Psi^{j_{k+1}}(\tau_{k+1}+0))|
^{2}=w(j_{k+1}/j_{k}),
\label{2.4.3}
\end{equation}
\begin{equation}
w_{j_{k}\gets\ldots\gets j_{k+m}}=w_{j_{k+m}\gets\ldots\gets j
_{k}}.
\label{2.4.4}
\end{equation}

The probabilities satisfy the equations
\begin{equation}
\sum_{j_{k+1}}w(j_{k+1}/j_{k})=\sum_{j_{k}}
w(j_{k+1}/j_{k})=1.
\label{2.4.5}
\end{equation}

We obtain for $t=\tau$ by Bayes formula, under the condition
$w(j_{k})={\rm const}$,
\begin{equation}
w(j_{k}/j_{k+1})=w(j_{k+1}/j_{k}),
\label{2.4.6}
\end{equation}
which coincides with eq.(\ref{2.4.3}).

\subsection{Nonorientability of standard indeterministic
quantum dynamics}

Summing up the results of this section, we conclude that
the dynamics of standard indeterministic quantum theory is
nonorientable and, by the same token, does not fix the
orientation of physical time.

\section{Dynamics of indeterministic quantum gravity}

As in standard quantum theory, in indeterministic quantum
gravity indeterminism originates from quantum jumps. But
the origin of the jumps in the latter theory differs radically
from that in the former one.

A quantum jump is the reduction of a state vector to one of
its components. In standard quantum theory, the cause of the
jump is coherence breaking between the components. In
indeterministic quantum gravity, the cause is energy difference
between the components, the difference occurring at a crossing of
energy levels.

According to the paper [2], a jump occurs at the tangency of
two levels. But level tangency imposes too severe constraints
on the occurrence of the jump. Here we introduce a scheme in
which the jump occurs at a simple crossing of two levels.

\subsection{Level crossing}

Let $\tau=0$ be the point of a crossing of levels $l=1,2$;
$P_{1\tau},P_
{2\tau}$ be the projectors for the corresponding
states in a neighborhood of the point:
\begin{equation}
P_{l\tau}\leftrightarrow \omega_{ml\tau}=
(\Psi_{l\tau},\cdot\Psi_{l\tau})
\label{3.1.1}
\end{equation}
($m$ stands for matter), and
\begin{equation}
P_{\tau}=P_{1\tau}+P_{2\tau}.
\label{3.1.2}
\end{equation}
The part of the Hamiltonian $H_{\tau}$ related to the
two levels is a projected Hamiltonian
\begin{equation}
H_{\tau}^{\rm proj}=P_{\tau}H_{\tau}P_{\tau}=
\epsilon_{1\tau}P_{1\tau}+\epsilon_{2\tau}P_
{2\tau}
\label{3.1.3}
\end{equation}
$(H_{\tau}^{\rm proj}\;{\rm is}\;\tilde H_{t}\;
{\rm in}\;[2])$. The metric tensor is
\begin{equation}
g=d\tau\otimes d\tau-h_{\tau}
\label{3.1.4}
\end{equation}
$(h_{\tau}\;{\rm is}\;\tilde g_{t}\;{\rm in}\;[2])$.

We have
\begin{equation}
H_{\tau}^{\rm proj}=H^{\rm proj}[h_{\tau},\dot h_{\tau}],
\label{3.1.5}
\end{equation}
where dot denotes the derivative with respect to the pretime
$\tau$,
\begin{equation}
H_{0}^{\rm proj}=\epsilon_{0}P_{0}=H^{\rm proj}
[h_{0},\dot h_{0}],\quad \epsilon_{0}=\epsilon_{10}=
\epsilon_{20}.
\label{3.1.6}
\end{equation}

\subsection{Creation projector and creation state}

We have in the first order in $\tau$
\begin{equation}
H_{\tau}^{\rm proj}=H_{0}^{\rm proj}+\dot H_{0}^{\rm proj}\tau
=H_{0}^{\rm proj}+\dot H^{\rm proj}[h_{0},\dot h_{0},\ddot h
_{0}]\tau.
\label{3.2.1}
\end{equation}
Furthermore,
\begin{equation}
\ddot h_{0}=\ddot h[h_{0},\dot h_{0},P^{\rm creat}],
\label{3.2.2}
\end{equation}
where $P^{\rm creat}$ is a one-dimensional projector which
creates $\ddot h_{0}$ and, by the same token, the Hamiltonian
$H_{\tau}^{\rm proj}$ eq.(\ref{3.2.1}). This creation projector
satisfies
\begin{equation}
P^{\rm creat}P_{0}=P^{\rm creat}
\label{3.2.3}
\end{equation}
and corresponds to a creation state $\omega_{m}^{\rm creat}$
belonging to a state subspace determined by $P_{0}$. For
the sake of brevity, we write
\begin{equation}
H_{\tau}^{\rm proj}=H_{0}^{\rm proj}+v\tau,
\label{3.2.4}
\end{equation}
\begin{equation}
v=\dot H^{\rm proj}[h_{0},\dot h_{0},\ddot h[h_{0},
\dot h_{0},P^{\rm creat}]].
\label{3.2.5}
\end{equation}

\subsection{Diagonal Hamiltonian}

The diagonalization of the Hamiltonian $H_{\tau}^{\rm proj}$
eq.(\ref{3.2.4}) gives
\begin{equation}
H_{\tau}^{\rm proj}=(\epsilon_{0}+\epsilon_{\tau}^{+})
P^{+}
+(\epsilon_{0}+\epsilon_{\tau}^{-})P^{-},
\label{3.3.1}
\end{equation}
where
\begin{equation}
\epsilon_{\tau}^{\pm}=\tau\frac{v_{11}+
v_{22}}{2}\pm |\tau|\sqrt{\frac{(v_{11}-v_{22})^{2}}{4}
+|v_{12}|^{2}},
\label{3.3.2}
\end{equation}
\begin{equation}
P^{\pm}\leftrightarrow\omega_{m}^{\pm}=(\Psi^{\pm},\cdot
\Psi^{\pm}),
\label{3.3.3}
\end{equation}
\begin{equation}
\Psi^{+}=e^{{\rm i}\beta}\cos\vartheta\:
\Psi_{1}+\sin\vartheta
\:\Psi_{2},\quad\Psi^{-}=-e^{{\rm i}\beta}\sin\vartheta\:
\Psi_{1}
+\cos\vartheta\:\Psi_{2},
\label{3.3.4}
\end{equation}
\begin{equation}
e^{{\rm i}\beta}=\frac{\tau}{|\tau|}\frac{v_{12}}{|v_{12}|},
\label{3.3.5}
\end{equation}
\begin{equation}
\tan\vartheta=\frac{|v_{12}|}{(\tau/|\tau|)(v_{11}-v_{22})/2
+\sqrt{(v_{11}-v_{22})^{2}/4+|v_{12}|^{2}}},
\label{3.3.6}
\end{equation}
\begin{equation}
\cot\vartheta=\frac{|v_{12}|}{(-\tau/|\tau|)(v_{11}-v_{22})
/2
+\sqrt{(v_{11}-v_{22})^{2}/4+|v_{12}|^{2}}},
\label{3.3.7}
\end{equation}
\begin{equation}
v_{ll'}=(\Psi_{l},v\Psi_{l'}),
\label{3.3.8}
\end{equation}
and $\{\Psi_{1},\Psi_{2}\}$ is a basis in the
two-dimensional Hilbert subspace ${\cal H}_{0}^{(2)}$
determined by $P_{0}$.

We have
\begin{equation}
\epsilon_{-\tau}^{\pm}=-\epsilon_{\tau}^{\mp},
\label{3.3.9}
\end{equation}
\begin{equation}
\tau\to -\tau\Rightarrow e^{{\rm i}\beta}\to-e^{{\rm i}\beta},
\tan\vartheta\leftrightarrow \cot\vartheta,\sin\vartheta
\leftrightarrow\cos\vartheta,\Psi^{+}\leftrightarrow\Psi^{-},
P^{+}\leftrightarrow P^{-}.
\label{3.3.10}
\end{equation}

\subsection{Germ projector, germ state, and germ process}

A germ projector $P^{\rm germ}$ is one of the two
projectors
$P^{\pm}$ eq.(\ref{3.3.3}); it gives rise to a germ
process ${\cal P}^{\rm germ}$---a process in a proximity
of the point $\tau=0$; ${\cal P}_{(0,\theta)}^{\rm germ\;right}$
and
${\cal P}_{(-\theta,0)}^{\rm germ\;left}$ are defined by

(i) ${\cal P}^{\rm germ}$ is deterministic;

(ii) $\lim_{\tau\to+0}{\cal P}_{(0,\theta)}^{\rm germ\;right}
(\tau)=\omega_{m}^{\rm germ\;right}\leftrightarrow P
^{\rm germ\;right}$;

(iii) $\lim_{\tau\to-0}{\cal P}_{(-\theta,0)}^{\rm germ\; left}
(\tau)=\omega_{m}^{\rm germ\;left}\leftrightarrow P
^{\rm germ\;left}$.

For $\Psi\in{\cal H}_{0}^{(2)}$ we put
\begin{equation}
\Psi=e^{{\rm i}\alpha}\cos\varphi\:\Psi_{1}+\sin
\varphi\:\Psi_{2},
\label{3.4.1}
\end{equation}
so that
\begin{equation}
P^{\rm creat}\leftrightarrow(\alpha,\varphi).
\label{3.4.2}
\end{equation}
The operator $v$ eq.(\ref{3.2.5}) is a function of
$(\alpha,\varphi)$, so that $\beta,\theta$
eqs.(\ref{3.3.5}),(\ref{3.3.6}),(\ref{3.3.7})
are such functions as well,
\begin{equation}
(\alpha,\varphi)\to(\beta,\theta).
\label{3.4.3}
\end{equation}
We assume that there exist the inverse functions,
\begin{equation}
(\beta,\theta)\to(\alpha,\varphi),
\label{3.4.4}
\end{equation}
so that there exists a bijection
\begin{equation}
(\alpha,\varphi)\leftrightarrow(\beta,\theta).
\label{3.4.5}
\end{equation}
As
\begin{equation}
P^{+}+P^{-}=P_{0},
\label{3.4.6}
\end{equation}
so that
\begin{equation}
P^{+}\leftrightarrow P^{-}
\label{3.4.7}
\end{equation}
and
\begin{equation}
P^{\rm germ}\to\{P^{+},P^{-}\}\leftrightarrow
(\beta,\theta),
\label{3.4.8}
\end{equation}
we have
\begin{equation}
{\cal P}^{\rm germ}\leftrightarrow P^{\rm germ}\to(\beta,
\theta)
\leftrightarrow (\alpha,\varphi)\leftrightarrow P^{\rm
creat}.
\label{3.4.9}
\end{equation}
Thus
\begin{equation}
{\cal P}^{\rm germ}\to P^{\rm creat}.
\label{3.4.10}
\end{equation}

\subsection{Regular crossing}

Let for $\tau<0$
\begin{equation}
\omega_{m}^{\rm creat\:left}=\omega_{m}^{\rm germ\:left}
\label{3.5.1}
\end{equation}
hold. Then it is natural to put for $\tau>0$
\begin{equation}
\omega_{m}^{\rm creat\:right}=\omega_{m}^{\rm creat\:left}
\equiv\omega_{m}^{\rm creat}
\label{3.5.2}
\end{equation}
and, in view of eqs.(\ref{3.3.9}),(\ref{3.3.10}),
\begin{equation}
\omega_{m}^{\rm germ\:right}=\omega_{m}^{\rm germ\:left}.
\label{3.5.3}
\end{equation}
Thus, there exists a germ process ${\cal P}_{(-\theta,\theta)}
^{\rm germ}$, such that ${\cal P}_{(-\theta,0)}^{\rm germ\:
left}$ and
${\cal P}_{(0,\theta)}^{\rm germ\:right}$ are its restrictions,
\begin{equation}
\lim_{\tau\to-0}{\cal P}^{\rm germ\:left}(\tau)=\lim_{\tau\to+0}
{\cal P}^{\rm germ\:right}(\tau)={\cal P}_{(-\theta,\theta)}
^{\rm germ}(0)=
\omega_{m}^{\rm creat},
\label{3.5.4}
\end{equation}
and there is no jump. The point $\tau=0$ and the process
${\cal P}_{(-\theta,\theta)}^{\rm germ}$ are deterministic.

\subsection{Singular crossing and quantum jump}

Now let
\begin{equation}
\omega_{m}^{\rm creat\:left}\ne\omega_{m}^{\rm germ\:left}.
\label{3.6.1}
\end{equation}
There is no possibility for a continuous process
${\cal P}_{(-\theta,\theta)}^{\rm germ}$. Since $\omega_{m0}$
is not
determined by the process ${\cal P}_{(-\theta,0)}^{\rm germ\:
left}$,
we put
\begin{equation}
\omega_{m0}=\lim_{\tau\to-0}\omega_{m\tau}=\omega_{m}^{\rm
germ\:left}.
\label{3.6.2}
\end{equation}
Furthermore, it is natural to put
\begin{equation}
\omega_{m}^{\rm creat\:right}=\omega_{m0},
\label{3.6.3}
\end{equation}
so that
\begin{equation}
\omega_{m}^{\rm creat\:right}=\omega_{m}^{\rm germ\:left}=
\lim_{\tau\to-0}\omega_{m\tau}=\omega_{m-0}.
\label{3.6.4}
\end{equation}

We have, by
eqs.(\ref{3.4.2}),(\ref{3.4.3}),(\ref{3.4.8}),(\ref{3.6.4}),
a quantum jump
\begin{equation}
P_{-0}^{\rm left}\leftrightarrow \omega_{m-0}
\stackrel{\rm jump}{\longrightarrow}\omega_{m+0}^{l}
\leftrightarrow
P_{+0}^{{\rm right}\:l},\quad l=\pm,
\label{3.6.5}
\end{equation}
with a transition probabilities related to it
\begin{equation}
w({\cal P}_{(0,\theta)}^{{\rm germ\:right}\:l}
/{\cal P}_{(-\theta,0)}
^{\rm germ\:left})={\rm Tr}\{P_{+0}^{{\rm right}\:l}
P_{-0}^{\rm left}\},\quad l=\pm.
\label{3.6.6}
\end{equation}

In the case of a regular crossing, eq.(\ref{3.6.6}) gives
$w=1$ or 0; this case is an idealized limiting one.

Thus, a singular crossing gives rise to a quantum jump.

\subsection{Orientability of dynamics and arrow of time}

A point which corresponds to a singular crossing is
indeterministic. We have for the cardinal numbers related to
it
\begin{equation}
{\rm card^{future\:\alpha}=card^{right\:\alpha}=2>1=
card^{left\:\alpha}=card^{past\:\alpha}},\quad \alpha=l=\pm.
\label{3.7.1}
\end{equation}
Thus the dynamics of indeterministic quantum
gravity is orientable;
it determines the arrow of time given by
\begin{equation}
t=\tau.
\label{3.7.2}
\end{equation}

We find for the probabilities of subsection 1.8
\begin{equation}
w(f/i)={\rm Tr}\{P_{+0}^{{\rm right}\:l}
P_{-0}^{\rm left}\},\quad i=1,\quad f=l=\pm,
\label{3.7.3}
\end{equation}
\begin{equation}
\sum_{f}w(f/i)={\rm Tr}\{P_{0}P
_{-0}^{\rm left}\}=
{\rm Tr}\{P_{-0}^{\rm left}\}=1,
\label{3.7.4}
\end{equation}
\begin{equation}
\sum_{i}w(f/i)=w(f/1)\leq 1,
\label{3.7.5}
\end{equation}
\begin{equation}
w(i/f)=1,
\label{3.7.6}
\end{equation}
so that
\begin{equation}
w(f/i)\leq w(i/f)=1.
\label{3.7.7}
\end{equation}

\subsection{Nonpredeterminability of the future and
reconstructibility of the past}

The dynamics developed is indeterministic, therefore the
future
is not predeterminate and may be forecasted only on a
probabilistic level. On the other hand, in view of
eqs.(\ref{3.4.10}),(\ref{3.6.4}), we have
\begin{equation}
{\cal P}_{(0,\theta)}^{\rm germ\:future}\to \omega_{m}
^{\rm create\:right}=\omega_{m}^{\rm germ\:left}
\leftrightarrow
{\cal P}_{(-\theta,0)}^{\rm germ\:past},
\label{3.8.1}
\end{equation}
so that
\begin{equation}
{\cal P}_{(0,\theta)}^{\rm germ\:future}\to {\cal P}
_{(-\theta,0)}
^{\rm germ\:past}.
\label{3.8.2}
\end{equation}
Thus, the past is reconstructible uniquely.

\end{document}